\documentclass[aps,pra,twocolumn,floatfix,10pt,nofootinbib]{revtex4-1} 

\usepackage{amsmath,amsfonts,amssymb,bbm,times}
\usepackage{dsfont}
\usepackage{graphicx}
\usepackage[usenames,dvipsnames]{xcolor}   
\usepackage{ulem}                          

\usepackage{hyperref}

\DeclareMathOperator{\tr}{tr}

\hypersetup{colorlinks=true,
						linkcolor=MidnightBlue,        
						citecolor=MidnightBlue,        
						filecolor=magenta,      			 
						urlcolor=MidnightBlue          
}

\begin{document}

\bibliographystyle{apsrev}
\newtheorem{example}{Example}

\newtheorem{remark}{Remark}
\newtheorem{problem}{Problem}
\newtheorem{theorem}{Theorem}
\newtheorem{corollary}{Corollary}
\newtheorem{definition}{Definition}
\newtheorem{proposition}{Proposition}
\newtheorem{lemma}{Lemma}
\newcommand{\proofend}{\hfill\fbox\\\medskip }
\newcommand{\proof}[1]{{\bf{Proof.}} #1 $\proofend$}
\newcommand{\nn}{{\mathbbm{N}}}
\newcommand{\rr}{{\mathbbm{R}}}
\newcommand{\cc}{{\mathbbm{C}}}
\newcommand{\zz}{{\mathbbm{Z}}}
\newcommand{\je}{\ensuremath{\heartsuit}}
\newcommand{\jd}{\ensuremath{\clubsuit}}
\newcommand{\id}{{\mathbbm{1}}}
\renewcommand{\vec}[1]{\boldsymbol{#1}}
\newcommand{\me}{\mathrm{e}}
\newcommand{\mi}{\mathrm{i}}
\newcommand{\md}{\mathrm{d}}
\newcommand{\sg}{\text{sgn}}
\newcommand{\mc}[1]{\textcolor{blue}{[M.C.: #1]}}

\def\>{\rangle}
\def\<{\langle}
\def\({\left(}
\def\){\right)}

\newcommand{\ket}[1]{\left|#1\right>}
\newcommand{\bra}[1]{\left<#1\right|}
\newcommand{\braket}[2]{\<#1|#2\>}
\newcommand{\ketbra}[2]{\left|#1\right>\!\left<#2\right|}
\newcommand{\proj}[1]{|#1\>\!\<#1|}
\newcommand{\avg}[1]{\< #1 \>}

\renewcommand{\tensor}{\otimes}

\newcommand{\einfuegen}[1]{\textcolor{PineGreen}{#1}}
\newcommand{\streichen}[1]{\textcolor{red}{\sout{#1}}}
\newcommand{\todo}[1]{\textcolor{blue}{(ToDo: #1)}}
\newcommand{\transpose}[1]{{#1}^t}

\newcommand{\magenta}[1]{\textcolor{red}{#1}}
\newcommand{\mbp}[1]{{\color{blue} #1}}

\newcommand{\om}[1]{{\color{black} #1}}

\title{Practical Entanglement Estimation for Spin-System Quantum Simulators}

\author{O. Marty, M. Cramer and M.B. Plenio}
\affiliation{Institut f\"ur Theoretische Physik \& IQST, Albert-Einstein-Allee 11, Universit\"at Ulm, Germany}

\begin{abstract}
We present practical methods to measure entanglement for quantum simulators that can be realized with trapped
ions, cold atoms, and superconducting qubits. Focussing on long- and short-range Ising-type Hamiltonians, 
we introduce schemes that are applicable under realistic experimental conditions including mixedness due to, \om{e.g.,\ noise or temperature}. 
In particular, we identify a single observable whose expectation value serves as a 
lower bound to entanglement and which may be obtained by a simple quantum circuit. As such circuits are not 
(yet) available for every platform, we investigate the performance of routinely measured observables as 
quantitative entanglement witnesses. Possible applications include experimental studies of entanglement 
scaling in critical systems and the reliable benchmarking of quantum simulators.

\end{abstract}

\maketitle

\date{\today}

\section{Introduction}

Harnessing the potential of well-controlled experimental platforms, quantum simulators 
have recently emerged as analogue devices to study paradigmatic condensed-matter models
\cite{JohnsonClarkJaksch14}. To date, a considerable variety of devices have been
proposed and partially realised to serve the central aim in this field, the preparation
and control of quantum states with a number of constituents that is beyond the reach of classical 
simulations \cite{GeorgescuAshhabNori14}. For the demonstration of genuinely quantum 
features of these simulators, it is thus of considerable interest to find methods which quantify 
entanglement and, if possible, relate the findings to classical simulatability. For pure 
states, the bi-partite block entanglement is a direct figure of merit for
the resources required when simulating many-body systems with numerical methods such as the 
density-matrix renormalization group~\cite{SchuchWolfVerstraeteCirac08,Schollwoeck11,EisertCramerPlenio08}. 
One way to obtain the entanglement contained in a state in the laboratory would be to perform 
full quantum state tomography \cite{Tomography} and to compute the entanglement of the reconstructed 
state. However, this is not only impractical due to the exponential resources required---the 
proverbial {\it curse of dimensionality}---but for many reconstruction schemes it may also 
lead to a systematic overestimation of the true entanglement content~\cite{Schwemmer2015}. An 
experimentally feasible and rigorous alternative is to instead rely on lower bounds which may 
be obtained directly from measured observables 
\cite{Horodecki99,Brandao2005,AudenaertPlenioNJP06,EisertBrandaoAudenaert,GuehneReimpellWerner1,GuehneReimpellWerner2} 
and such lower bounds to the entanglement should (i) rely only on a few observables in order to
avoid the curse of dimensionality, (ii) avoid assumptions on the state in the laboratory (such 
as, e.g., symmetries, temperature or an underlying Hamiltonian), and (iii) should be applicable 
to the experimentally relevant setting of mixed states. Indeed, as has already been demonstrated, 
(i)-(iii) may be met and entanglement may be quantified from significantly less observables than 
are required for the knowledge of the full state: E.g., collective observables are capable to detect \cite{WiesniakVedralBrukner05,TothKnappGuehneBriegel07,KrammerKampermann2009} and quantify 
\cite{CramerPlenioWunderlich11,Marty2014,Cramer2013,LueckePeiseEtal14} entanglement. Note that 
extending (ii) also to observables, is known as device-independent entanglement quantification,
for which there does not even need to be a quantum description of the employed measurement device, 
see~\cite{Moroder2013} and references therein. Here, however, we will assume that the relevant 
observables are actually those that are measured.

We construct and analyze lower bounds to the bi-partite entanglement of states arising in the 
quantum simulation of a variety of spin models such as
\begin{equation}
    \label{eq:transversefieldising}
    \hat H = \sum_{i,j=1 }^N J_{i,j} \hat\sigma_z^i\hat\sigma_z^j + B \sum_{i=1}^N \hat\sigma_x^i,
\end{equation}
which have recently been implemented in experiments with trapped ions~\cite{Friedenauer08,Kim2010,IslamEdwardsKimEtal11,JurcevicLanyonHaukeEtal14,Richerme2014},
superconducting qubits~\cite{SuperCondQubits}, and ultra-cold atoms~\cite{SimonBakrMaTai11,Meinert2013}. 
We will consider ground states and their quasi-adiabatic dynamical preparation employing realistic 
noise models, including decoherence-induced mixedness.

Our aim is to quantify bi-partite block entanglement of one part of the chain vs.\ the rest relying 
only on measurements of certain observables $\hat{C}_i$. Denoting experimentally obtained expectation 
values of these observables by $c_i$ \cite{errorFootnote}, we are thus interested in
\begin{equation}
    \label{the main optimization}
    E_{\text{min}}[\{\hat{C}_i\},\{c_i\}]=\min_{\hat\varrho}\left\{E(\hat\varrho)\,\big|\,\text{tr}[\hat{C}_i\hat\varrho]=c_i\right\},
\end{equation}
i.e., we consider the minimal amount of entanglement that is consistent with the obtained measurements 
$c_i$. Here, $E$ is the entanglement measure of choice and the minimization is taken over all density 
matrices $\hat{\varrho}$. As such, we follow the programme initiated in Refs.~\cite{Horodecki99,Brandao2005,AudenaertPlenioNJP06,EisertBrandaoAudenaert,GuehneReimpellWerner1,GuehneReimpellWerner2}.
Note that no assumption on the state in the laboratory enters our considerations. While we will present 
tailored lower bounds to $E_{\text{min}}$ that work particularly well---in some cases even providing 
$E_{\text{min}}$ exactly---for certain classes of states, we stress that all bounds presented in this 
work are valid for {\it arbitrary} states -- pure or mixed.

For systems governed by Hamiltonians as in Eq.~\eqref{eq:transversefieldising}, we identify a single 
key quantity in order to obtain lower bounds on $E_{\text{min}}$. That is, it turns out that a single 
observable $\hat{C}$ constitutes a common quantitative witness and, in fact, for large classes of states  
determines not only a lower bound but the entanglement of $\hat \varrho$ itself. We show how this 
witness may be measured directly by employing a simple quantum circuit. If such a circuit is available, 
entanglement may thus be quantified for systems consisting of an arbitrary number of spins. If it is not 
available, the above observation still allows us to transform the numerical minimization in 
Eq.~\eqref{the main optimization} into the problem of computing the smallest eigenvalue of a sparse matrix
and thus obtain results for more than $20$ spins (and in principle many more using DMRG methods \cite{Schollwoeck11}).
With recent implementations of models as in Eq.~\eqref{eq:transversefieldising} in mind, we thus introduce 
schemes for practical and rigorous experimental entanglement estimation  using only a few readily available 
observables and without relying on any assumptions on the state in laboratory.

Throughout, we will use the logarithmic negativity \cite{PlenioVirmani07} as our bipartite entanglement measure
and consider the bipartition $\{1,\dots,\frac{N}{2}\}|\{\frac{N}{2}+1,\dots,N\}$, assuming $N$ to be even. 
The logarithmic negativity is a full entanglement monotone for mixed states \cite{LogNeg}, an upper bound 
to the distillable entanglement \cite{distBound}, and has an operational interpretation \cite{LogNegInterpretation}. 
It reduces to the \om{R\'enyi entanglement entropy with R\'enyi index $1/2$} on pure states, which, e.g., distinguishes topologically 
ordered phases (as do all the R\'enyi entanglement entropies \cite{FlammiaHammaEtal09}). In a setting involving 
mixed states, a topological contribution to the logarithmic negativity of the toric code model has been established 
in \cite{Castelnovo13}.

\section{Preliminaries}
\label{section:preliminaries}
We start by introducing the relevant quantities.
The logarithmic negativity is defined as
\begin{equation}
    E_{\text{l.n.}}(\hat\varrho) = \log \|\hat\varrho^\Gamma\|_1,
\end{equation}
where $ \hat\varrho^{\Gamma}$ is the partial transpose of $\hat \varrho$
with respect to the chosen bipartition $A|B$ (here, $\{1,\dots,\frac{N}{2}\}|\{\frac{N}{2}+1,\dots,N\}$) and
\begin{equation}
    \label{eq:tracenorm}
    \|\hat X \|_1 = \tr |\hat X|  = \max\bigl\{\tr(\hat C \hat X) \,\big|\, -\id \leq \hat C \leq \id \bigr\}.
\end{equation}
is the trace norm. By its variational form we have that for any observable with $-\id \leq \hat C^\Gamma \leq \id$
\begin{equation}
    \label{eq:quantitativewitness}
    E_{\text{l.n.}}(\hat\varrho) \geq \log \langle\hat C\rangle_{\!\hat\varrho}.
\end{equation}
Any observable $\hat C$ with this property thus serves as a
quantitative entanglement witness as it not only witnesses entanglement
but indeed provides a lower bound.
As an important example  for such a quantitative witness consider the un-normalized
maximally entangled state
\begin{equation}
    \ket{\Phi}=2^{N/4}\bigotimes_{i=1}^{N/2}\ket{\phi}_{i,N+1-i},\;\;
    \ket{\phi}_{i,j}=\frac{|00\rangle+|11\rangle}{\sqrt{2}},
\end{equation}
which fulfils $-\id\le\bigl(\hat U\ketbra{\Phi}{\Phi}\hat U^\dagger\bigr)^\Gamma\le \id$
for any unitary $\hat U=\hat V\otimes\hat W$.
Hence, for any state $\hat\varrho$
\begin{equation}
    \label{eq:variationallogneg}
    E_{\text{l.n.}}(\hat\varrho) \geq \log \max_{\hat U=\hat V\otimes\hat W}\bigl\langle\hat U\ketbra{\Phi}{\Phi}\hat U^\dagger\bigr\rangle_{\!\hat\varrho}.
\end{equation}
The significance of the quantitative witness $\hat U\ketbra{\Phi}{\Phi}\hat U^\dagger$ becomes 
clear when considering pure states: For a given pure state, consider its Schmidt decomposition 
$|\psi\rangle=\sum_s\psi_s|a_s\rangle|b_s\rangle$ and let $\hat U=\hat V\otimes\hat W$ be the 
unitary that takes $|\Phi\rangle$ to $\sum_s|a_s\rangle|b_s\rangle$. Then $\langle\psi|\hat{U}|\Phi\rangle
=\|(|\psi\rangle\langle\psi|)^\Gamma\|_1^{1/2}$ and thus Eq.~(\ref{eq:variationallogneg}) becomes 
an equality.

\begin{figure*}[t]
\includegraphics[width=0.95\textwidth]{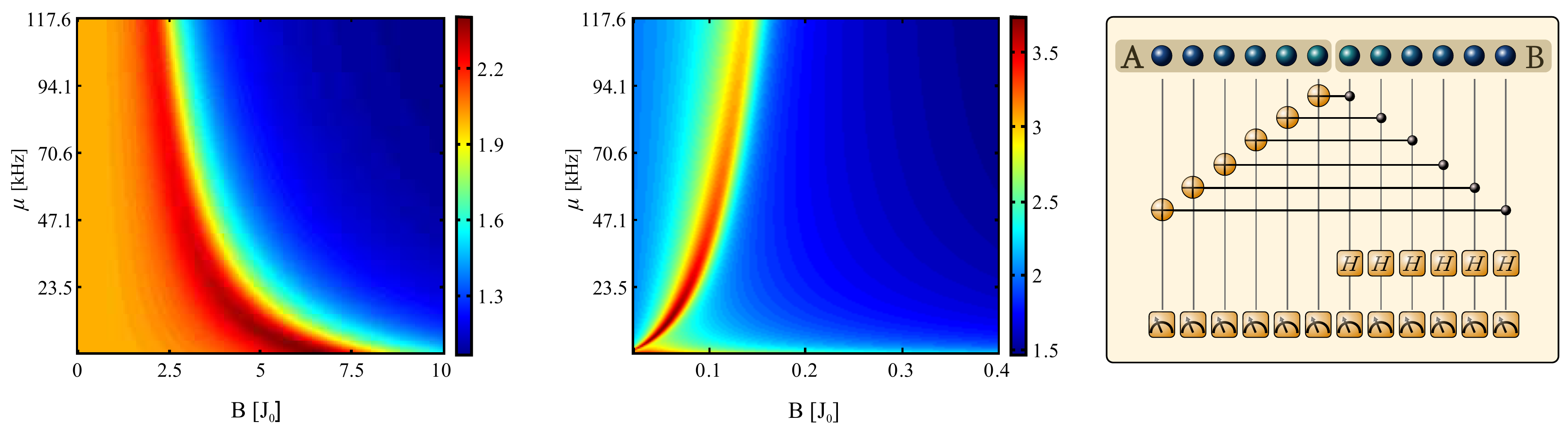}
\caption{\label{fig:swapandprotocol}%
Expectations of $|\Phi\rangle\langle\Phi|$ for the ground state of the ferromagnetic (left) and of $|\Phi^\prime\rangle\langle\Phi^\prime|$
for the ground state of the antiferromagnetic (middle) long-range Ising Hamiltonian in Eq.~\eqref{eq:transversefieldising} for realistic couplings \cite{SimulationParameters}
as in Eq.~\eqref{eq:iontrapisingcouplings} with $N=20$ and $B$ in units of $J_0 = \sum_i |J_{i,i+1}|/(N-1)$. The coupling range is determined by the detuning parameter $\mu$. The expectations may be obtained via the circuit in Eq.~\eqref{eq:bellcircuit}, depicted on the right, and they coincide with the entanglement in the ground state, see Corollary~\ref{prop:isinggroundstateln}.}
\end{figure*}

While in general this requires the knowledge of the Schmidt vectors, we will see below that
for large classes of states, equality may be achieved for one particularly simple unitary.
This fact may be used to greatly simplify the optimization in Eq.~(\ref{the main optimization}). 
Furthermore, for these classes of states, $\langle\hat U\ketbra{\Phi}{\Phi}\hat U^\dagger\rangle_{\!\hat\varrho}$ 
may be obtained directly by applying a simple quantum circuit as in Fig.~\ref{fig:swapandprotocol} 
consisting of mutually commuting $N/2$ two-qubit controlled-not and $N/2$ single-qubit gates
and subsequently performing a projective measurement of $|0\rangle\langle 0|^{\otimes N}$ in the 
computational basis.

\section{Results}

The Ising model in Eq.~\eqref{eq:transversefieldising} has been realized on a variety of 
experimental platforms: Systems with tunable interactions are for example found in devices 
based on superconducting qubits \cite{SuperCondQubits}. Short-ranged couplings are encountered 
in experiments with ultra-cold atoms in optical lattices, see, e.g., Ref.~\cite{SimonBakrMaTai11}, 
in which nearest-neighbour interactions have been simulated. For ion-traps, the implementation 
of Eq.~\eqref{eq:transversefieldising} has been proposed theoretically \cite{PorrasCirac04} 
and realised experimentally \cite{Friedenauer08,Kim2010,IslamEdwardsKimEtal11}. Here, 
the basic form of $J_{i,j}$ is dictated by the properties of the trap and external laser fields:
For the scheme demonstrated in \cite{KimChangIslamKorenblit09}, Ising couplings are generated by 
two non-copropagating laser beams with frequencies $\omega_0 \pm \mu$ where $\omega_0$ denotes 
the energy splitting between the states defining a local spin, e.g.\ hyperfine clock states of 
$^{171}$Yb$^+$ \cite{IslamEdwardsKimEtal11,KimChangIslamKorenblit09,IslamSenkoCampellKorenblit12}.
If the beatnote detuning $\mu$ is sufficiently far from each (transversal) normal mode frequency 
$\omega_m$, in the usual rotating wave approximation and within the Lamb-Dicke regime this leads
to an effective Ising Hamiltonian with couplings given by
\begin{equation}
    \label{eq:iontrapisingcouplings}
    J_{i,j} = \Omega_i \Omega_j \frac{(\hbar k)^2}{4 M} \sum_m \frac{b_{i,m}b_{j,m}}{\mu^2 - \omega_m^2},
\end{equation}
where $\Omega_i$ is the Rabi frequency of the $i$th ion, $k$ the wave vector difference of the 
laser beams, $M$ the mass of the ions and $b_{i,m}$ denotes the transformation between the 
vibrational site excitations and the normal modes \cite{MarquetSchmidtKalerJames03}. The sum 
runs over all normal modes. The range of the interaction can be controlled by the detuning $\mu$ 
from infinite range if $\mu$ is close to the center of mass mode frequency $\omega_N$, where 
all the spins couple equally to the motional degrees of freedom, to dipole-dipole interactions 
for $\mu \gg \omega_N$ \cite{IslamSenkoCampellKorenblit12}. Alternatively the interaction range 
may be varied by changing the axial trap frequency $\nu_z$ \cite{SchachenmayerLanyonRoosDaley13}.
In between these two regimes the couplings are well approximated by an algebraic decay,
\begin{equation}
    \label{eq:algebraicdecayingcouplings}
    J_{i,j} = \frac{J}{|i-j|^p},
\end{equation}
with $0<p<3$. A transverse magnetic field may be introduced by an additional laser beam. 
Furthermore, ferromagnetic couplings may be obtained by choosing different detuning $\mu$ or 
by initializing the system in the highest excited state and following an adiabatic protocol 
\cite{IslamEdwardsKimEtal11}.

\subsection{The Quantum Circuit}
Our main result is that for ground states of a variety of spin Hamiltonians
the maximizing unitary in Eq.~\eqref{eq:variationallogneg} may be given explicitly:

\begin{corollary}
\label{prop:isinggroundstateln}
Let $\hat H$ a Hamiltonian as in Eq.~\eqref{eq:transversefieldising}  and suppose it has a 
non-degenerate ground state $\ket{\psi}$. Let the couplings be such that $J_{i,j} = J_{N+1-i,N+1-j}$. 
If the $\frac{N}{2} \times \frac{N}{2}$ matrix $\mathcal{J}$ with entries $\mathcal{J}_{i,j}=J_{i,N+1-j}$, 
$i,j = 1\ldots,\frac{N}{2}$, is negative semi-definite then
\begin{equation}
    E_{\text{l.n.}}(\ketbra{\psi}{\psi}) = \log_2 \tr[\ketbra{\Phi}{\Phi}\ketbra{\psi}{\psi}].
\end{equation}
If $\mathcal{J}$ is positive semi-definite then
\begin{equation}
    E_{\text{l.n.}}(\ketbra{\psi}{\psi}) = \log_2 \tr[\ketbra{\Phi^{\prime}}{\Phi^{\prime}}\ketbra{\psi}{\psi}],
\end{equation}
where $\ket{\Phi^\prime}=\hat{\sigma}_1^x\otimes\cdots\otimes\hat{\sigma}_{N/2}^x\ket{\Phi}$.
\end{corollary}

This is a corollary of a theorem allowing for even larger classes of Hamiltonians which we prove 
in the appendix, where we also show that the conditions of the corollary are met by couplings as 
in Eq.~\eqref{eq:algebraicdecayingcouplings}. \om{Furthermore, they are also met by the couplings as in Eq.~\eqref{eq:iontrapisingcouplings}
with, e.g., parameters as we choose them in the numerical examples}. Hence, the bipartite entanglement (between the left 
and right half of the chain as quantified in terms of the logarithmic negativity) of {\it any} state 
that is a non-degenerate ground state of a Hamiltonian as in Eq.~(\ref{eq:transversefieldising}) 
with couplings \om{fulfilling the hypotheses of} corollary \ref{prop:isinggroundstateln} is equal to the expectation value
of a simple (unnormalized) projector. What is more, this expectation value serves as a lower bound 
to the entanglement of {\it any} state (pure or mixed). One possibility to obtain this expectation 
value---so the overlap of the state in the laboratory with $|\Phi\rangle$, respectively 
$|\Phi^\prime\rangle$---is to apply a simple circuit and subsequently measuring the projector 
$|0\rangle\langle 0|^{\otimes N}$: For the ferromagnetic case, ($J<0$), we write 
$|\Phi\rangle=\hat{R}|0\rangle^{\otimes N}$, where
\begin{equation}
    \label{eq:bellcircuit}
    \hat R = \bigotimes_{i=1}^{N/2} \hat H_i \hat C_{i,N+1-i}
\end{equation}
and $\hat C_{i,j}$ denotes the controlled-not gate acting on spin $i$ (control) and $j$ (target) and
$\hat H_i$ the Hadamard gate acting on spin $i$. The antiferromagnetic case, ($J>0$), follows 
by additionally applying the transformation $\bigotimes_{i=1}^{N/2} \hat\sigma_x^i$ before the measurement.
Note that in ion trap experiments, spin polarization measurements along a particular axis are routinely 
performed by spin-dependent resonance fluorescence.

The logarithmic negativity of any state may thus be lower bounded by applying the circuit $\hat R$, 
which is depicted in Fig.~\ref{fig:swapandprotocol}. There, we also show numerical results for the 
thus obtained entanglement of the ground state of the Ising model in Eq.~\eqref{eq:transversefieldising} 
for realistic \cite{SimulationParameters} ferro- and antiferromagnetic couplings; cf. the phase diagram 
from the entanglement entropy in Ref.~\cite{KoffelLewensteinTagliacozzo12}.

Let us emphasize again that $\hat R$ neither depends on the couplings $J_{i,j}$ nor on the magnetic 
field $B$. Therefore, one does not require any knowledge about these parameters and the method 
is robust against an inexact implementation of the Hamiltonian \om{as long as the hypotheses of Corollary
\ref{prop:isinggroundstateln} hold. Also, as $\ketbra{\Phi}{\Phi}$ is a 
quantitative entanglement witness, the state in the laboratory $\hat \varrho$ does not need to be 
exactly in the ground state, it does not even need to be pure, in order for 
$\bigl\langle \hat{R}(|0\rangle\langle 0|)^{\otimes N}\hat{R}^\dagger\bigr \rangle_{\!\hat \varrho}$ 
to be a lower bound.}

\subsection{Other observables}
Although experimentally feasible (see e.g. Ref.~\cite{RiebeMonzKimEtal08} for the realization
of a CNOT gate in ion traps and Ref.~\cite{superconductingCNOT} for superconducting qubits),
other observables may be more accessible than the implementation of the circuit $\hat R$. To 
this end, we give lower bounds to Eq.~\eqref{the main optimization} in terms of arbitrary 
observables $\hat{C}_i$. Combining Eqs.~\eqref{the main optimization} and~\eqref{eq:variationallogneg}, 
we find that $E_{\text{min}}[\{\hat{C}_i\},\{c_i\}]$ is lower bounded by the logarithm of 
the solution to the semidefinite program (SDP)
\begin{equation}
    \label{opt new}
    \begin{split}
        \underset{w_i\in\rr}{\text{max}} &\;\, \sum_i w_i c_i\\
        \text{subject to} &\;\,  \sum_i w_i \hat C_i \leq \ketbra{\Phi}{\Phi}.
    \end{split}
\end{equation}
Considering this SDP instead of the original Eq.~\eqref{the main optimization} leads to a significant 
simplification of the optimization problem and standard SDP solvers like, e.g., SeDuMi \cite{Sturm99} 
may be used. Furthermore, the simplified SDP in Eq.~\eqref{opt new} is directly accessible to algorithms 
such as SDPNAL \cite{SaoSunToh10} or SDPAD \cite{WenGoldfarbYin10} intended for solving large-scale SDPs 
with (real) matrices of dimension more than $4000$ and number of constraints of the order of $10^6$. 
Therefore these algorithms may outperform standard interior point methods where they become too expensive
computationally. Note that the observables $\hat{C}_i$ in Eq.~\eqref{opt new} are entirely arbitrary 
and this scheme is thus sufficiently versatile to accommodate measurements of any experimental platform.

\begin{figure}[t!]
\includegraphics[width=0.42\textwidth,interpolate=false]{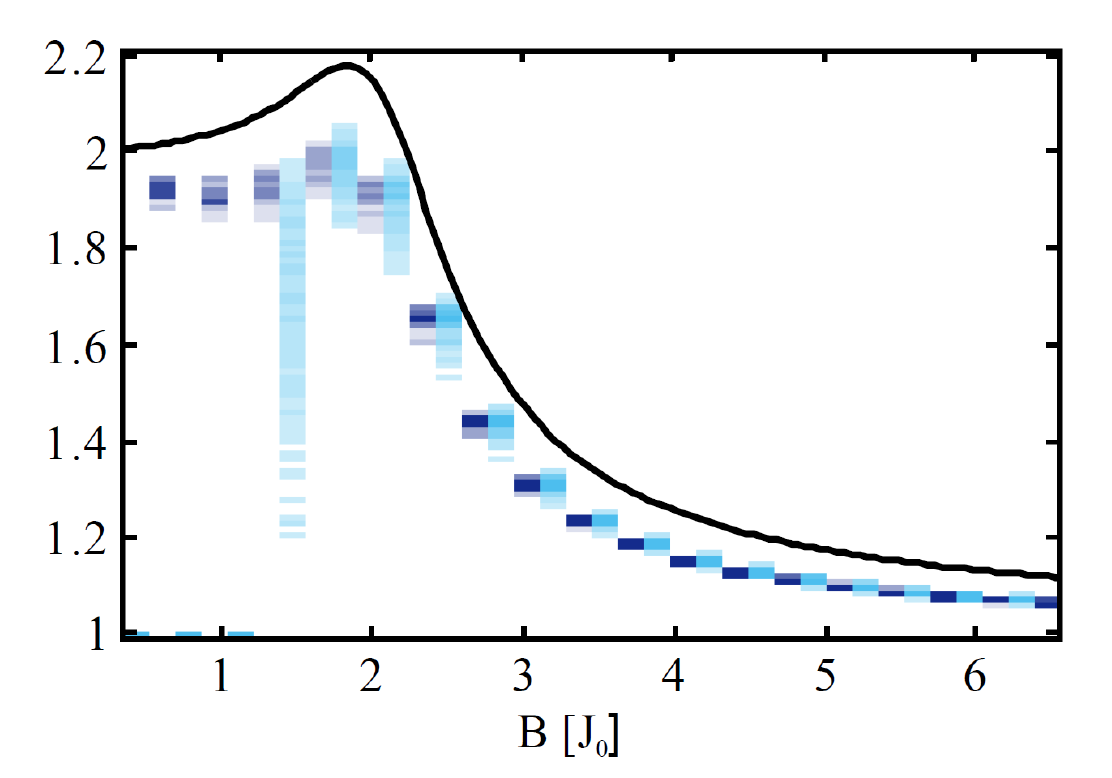}
\caption{\label{fig:boundfromNNwitness} Lower bounds to the entanglement of ground states
of the Hamiltonian in Eq.~\eqref{eq:transversefieldising} with $N=16$, $\mu = 117.6$ kHz, $J_0$ as in Fig.~\ref{fig:swapandprotocol}, and realistic couplings \cite{SimulationParameters} as in Eq.~\eqref{eq:iontrapisingcouplings}.
The black line shows the exact logarithmic negativity of the ground state.
Entanglement bounds are obtained by optimizing the quantitative witness in Eq.~\eqref{eq:NNwitness} over $w_1$. The couplings in the witness are as in Eq.~\eqref{eq:iontrapisingcouplings} but randomly perturbed by 2\% to mimic imprecise knowledge and shown are several random trials as density \om{with the optional operator $\bigotimes_{i=1}^N \hat\sigma_x^i$ in blue and without in cyan}.
}
\end{figure}

Motivated by the fact that if the ground state is separated from the first excited state by an energy 
gap, the Hamiltonian itself provides an entanglement witness \cite{hamiltonianFootnote}, we consider 
witnesses of the form
\begin{equation}
    \label{eq:NNwitness}
    \hat W = w_0 \id + \bigotimes_{i=1}^N \hat\sigma_x^i + w_1 \hat H,
\end{equation}
where we included the (optional, see Fig.\ \ref{fig:boundfromNNwitness}) operator $\bigotimes_{i=1}^N \hat\sigma_x^i$ to account for the small 
gap in the symmetry-broken phase. This further simplifies the optimization in Eq.~\eqref{opt new} as 
now we are considering only one observable---namely $\hat{W}$---and the number of optimization variables 
is reduced to one. Note that for $\hat{H}$ as in Eq.~\eqref{eq:transversefieldising}, the witness $\hat W$ 
consists of at most quadratically many observables (plus the single optional observable $\bigotimes_{i=1}^N 
\hat\sigma_x^i$) and hence its expectation value may in this sense be obtained efficiently: The experimental 
effort is reduced to obtaining the expectation value of the magnetization $\sum_i\hat{\sigma}^i_x$ and all 
pairs $\hat{\sigma}_z^i\hat{\sigma}_z^j$ for which $J_{i,j}$ is non-zero. In ion-trap and superconducting-qubit 
experiments such observables are routinely measured. For nearest-neighbour couplings as, e.g., the ultra-cold 
atoms experiment in Ref.~\cite{SimonBakrMaTai11}, this amounts to only linearly many observables, the correlators 
$\hat{\sigma}_z^i\hat{\sigma}_z^{i+1}$ may be obtained directly under a quantum-gas microscope \cite{SimonBakrMaTai11}, 
and the magnetization by a Fourier-transformation of the time-of-flight distribution. For the couplings 
one could either choose a theoretical prediction (for ion traps given in Eq.~\eqref{eq:iontrapisingcouplings}) 
or, if possible, measure them experimentally (see the methods used in Ref.~\cite{JurcevicLanyonHaukeEtal14} 
for ion traps). Then the SDP may be avoided completely by choosing $w_0$ as the smallest eigenvalue of
\begin{equation}
    \ketbra{\Phi}{\Phi} - \bigotimes_{i=1}^N \hat\sigma_x^i - w_1 \hat H
\end{equation}
as then the constraint $\hat{W}\le |\Phi\rangle\langle\Phi|$ is automatically fulfilled.
As this operator is a sparse matrix, standard eigenvalue solvers allow for system sizes
of more than 20 qubits. In fact, since $\ketbra{\Phi}{\Phi}$ possesses a representation 
as a matrix product operator of bond dimension four, DMRG algorithms may be used to obtain 
the smallest eigenvalue for much larger systems. In Fig.~\ref{fig:boundfromNNwitness} we 
show numerical results for the above procedure. Again, we do not put any assumptions on 
the state in the laboratory---the expectation $\langle\hat{W}\rangle_{\!\hat \varrho}$ is 
a lower bound to the entanglement of any state $\hat \varrho$ but, of course, we know that 
the bound will work particularly well for states that fall within the framework of 
Corollary~\ref{prop:isinggroundstateln}, i.e., ground states of Hamiltonians as in 
Eq.~\eqref{eq:transversefieldising} with couplings as in Eq.~\eqref{eq:algebraicdecayingcouplings} 
or as in Eq.~\eqref{eq:iontrapisingcouplings} with parameters as for all the numerical 
examples considered here.

\subsection{Quasi-adiabatic preparation and benchmarking}

\begin{figure*}[t]
\includegraphics[width=0.9\textwidth]{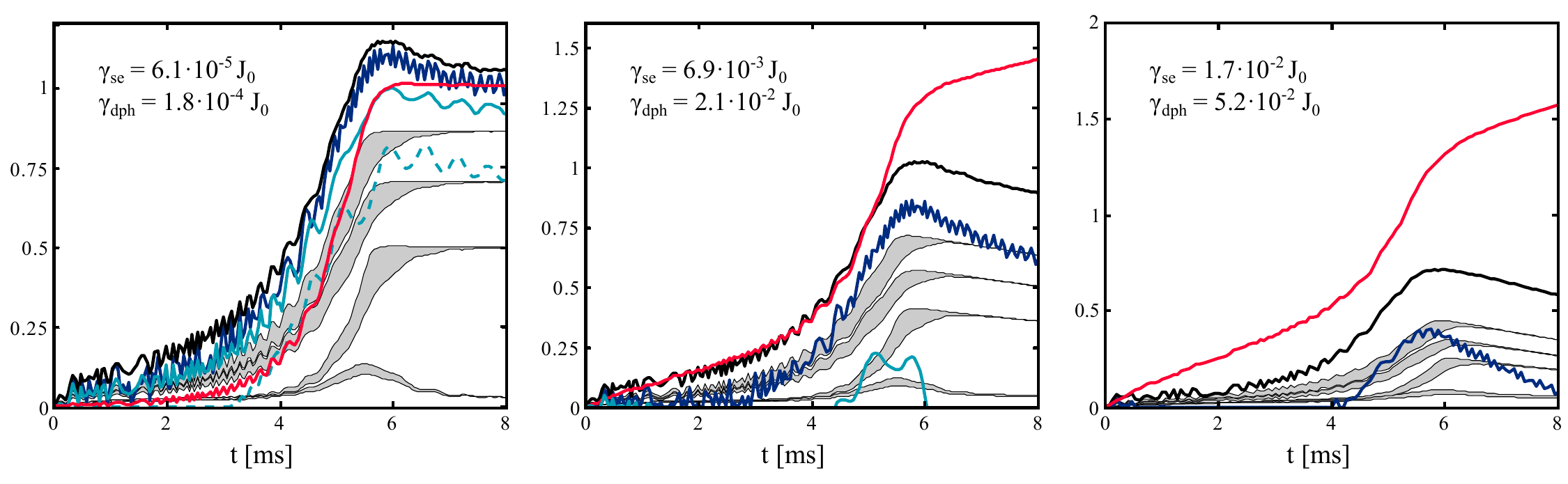}
\caption{\label{fig:lowvshighnoise} Quasi-adiabatic ramping of the magnetic field across the 
phase transition simulated using the model in Eq.~\eqref{lindblad} with Hamiltonian as in 
Eqs.~\eqref{eq:transversefieldising} and \eqref{eq:iontrapisingcouplings} for $N=8$ spins 
and parameters as in \cite{Fig3Parameters}, leading to a relatively long-ranged interaction 
$\sim |i-j|^{-0.3}$ and $J_0=2\pi \cdot 3.3$ kHz. The magnetic field is ramped according to 
$B = 1.1\cdot 2^{\kappa(t_0-t)/t_0} J_0$, where $\kappa = 2\pi\cdot 20$ kHz and $t_0 = 0.6$ ms.
Shown are the block entropy $S(\text{tr}_{1,\dots,N/2}[\hat{\varrho}(t)])$ (an entanglement 
measure if the state was pure) in red, the exact bi-partite entanglement of the simulated 
state $\hat{\varrho}(t)$ in black, lower bounds as obtained by the circuit $\hat{R}$ (blue), 
by the SDP in Eq.~\eqref{opt new} with observables $\hat \sigma_\alpha^i$, 
$\hat \sigma_\alpha^i\hat \sigma_\alpha^j$, $i,j=1,\dots,N$, $\alpha=x,y,z$, and 
$\hat \sigma_x^1\cdots\hat \sigma_x^N$ as input (solid cyan), and by optimizing the quantitative 
witness in Eq.~\eqref{eq:NNwitness} over $w_1$ (dashed cyan). Shown in grey are upper and lower 
bounds \cite{MPOerrorFootnote} to the MPO approximation error in Eq.~\eqref{MPOerror} ($D=1,2,3,4$ 
top to bottom). Note the monotonicity of the block entropy as opposed to the behaviour of the 
approximation error and the entanglement as quantified in terms of the logarithmic negativity.}
\end{figure*}

In non-equilibrium situations, quantum simulators of one-dimensional spin systems 
may outperform classical computers already for a moderate size of spins: As opposed 
to states in equilibrium, which typically have little entanglement (cf., {\it area 
laws} for ground and thermal states \cite{EisertCramerPlenio08,arealaws}), the 
entanglement generated in non-equilibrium situations may become large \cite{entanglementQuench}. 
Arguably the best numerical algorithms for the simulation of one-dimensional (non-)equilibrium 
quantum many-body systems are those based on matrix product states (MPS) and matrix 
product operators (MPO) \cite{Schollwoeck11,MPSO}. The resources required to treat 
such states numerically are directly related to their so-called {\it bond dimension}. 
For pure states, i.e. MPS, there is an intimate relation between the bond dimension and 
the entanglement content as quantified in terms of R\'enyi entanglement entropies \cite{SchuchWolfVerstraeteCirac08}.
For mixed states, i.e. MPO, this connection is far less clear. Indeed, an MPO may have 
a small bond dimension while at the same time have a large block entropy---the 
product operator $(\id/2)^{\otimes N}$ being the most striking example. In this sense, 
using pure-state entanglement measures (such as R\'enyi entanglement entropies) as 
benchmarks may lead to false conclusions because in experiments mixedness is unavoidable. 
We illustrate these relations by considering the quasi-adiabatic preparation of ground 
states of Ising Hamiltonians as commonly performed in ion-trap experiments~\cite{IslamEdwardsKimEtal11}: 
Initializing the system in a product state with all spins aligned parallel to the magnetic 
field, the field is reduced slowly (compared to the Ising interactions) until the desired 
$B$ is reached. In a realistic setting, such a protocol is prone to noise processes such as 
non-adiabaticity, spontaneous emission ($se$) and dephasing ($dph$), which are considered 
the main noise sources \cite{IslamEdwardsKimEtal11}. We model this by the commonly used 
Lindblad quantum master equation
\begin{equation}
    \label{lindblad}
    \frac{\md\hat\varrho(t) }{\md t}= -\mi \bigl[\hat H\!,\!\hat\varrho(t)\bigr] + \sum_{i,\alpha} \Bigl[\hat L_i^{\alpha\dag}\!\hat\varrho(t) \hat L_i^{\alpha} - \tfrac{1}{2} \bigl\{\!\hat L_i^{\alpha\dag} \hat L_i^{\alpha},\hat\varrho(t)\!\bigr\}\Bigr]
\end{equation}
with $\alpha=se, dph$ and $\hat L_i^{se} = \sqrt{\gamma_{se}} \hat\sigma_+^i$, 
$\hat L_i^{dph} = \sqrt{\gamma_{dph}} \hat\sigma_z^i$, and $\{\cdot,\cdot\}$ the 
anticommutator. Numerical results are summarized in Fig.~\ref{fig:lowvshighnoise} 
and the main conclusions are: The block entropy $S(\text{tr}_{1,\dots,N/2}[\hat{\varrho}(t)])$
(a measure of entanglement if the state was pure) increases with time for all 
noise-strengths while the true entanglement reaches a maximum after which it 
decreases in time. From the block entropy one would thus falsely conclude that the 
state becomes harder and harder to simulate while the error when approximating 
$\hat{\varrho}(t)$ by an MPO $\hat{\varrho}_D$ with bond dimension $D$,
\begin{equation}
    \label{MPOerror}
    \epsilon_D(t)=
    \min_{\hat{\varrho}_D}
    \|\hat{\varrho}(t)-\hat{\varrho}_D\|_F,
\end{equation}
reaches a maximum and then decreases in time \cite{MPOerrorFootnote} as does 
the entanglement. The exact mathematical connection between approximability by 
MPOs, entanglement, and other quantities such as, e.g., mutual information, remains 
an open question however.

\section{Summary and Outlook}

In the setting of quantum simulations of the transverse- field Ising model, we have 
introduced methods to estimate bi-partite block entanglement without putting any assumptions 
on the state in the laboratory. The principles presented here are applicable to, e.g., 
ion-trap, cold-gases, and superconducting-qubit implementations and we have focused 
on the ion-trap platform for specific examples. A lower bound to the entanglement 
is given by the overlap with a certain state, which may, e.g., be obtained by a simple 
quantum circuit and, for large classes of states, actually gives the entanglement exactly 
instead of just bounding it. As obtaining this overlap may, depending on the platform, 
may represent a considerable experimental challenge, we further investigated the 
performance of routinely performed measurements as means to estimate the entanglement. 
As we consider the benchmarking of quantum simulators as one possible application, we 
have compared the matrix-product-operator bond dimension, block entanglement, and block 
entropy for a quasi-adiabatic protocol preparing ground states of the transverse-field 
Ising model. It is our hope that this inspires work towards revealing the exact mathematical 
connection between the MPO approximation error and correlation measures and also towards 
schemes to estimate the former directly by measurements in a way similar to the schemes 
presented here for entanglement.

\section{Acknowledgements}
We thank A.\ Albrecht, M.\ Bruderer, N.\ Killoran, A.\ Lemmer, R.\ Puebla, A.\ Smirne and 
D.\ Tamascelli for comments and discussions, T.\ Baumgratz 
for discussions and providing the code of the variational compression method and the
bwGRiD project for computational resources. 
We acknowledge support from an Alexander von Humboldt-Professorship,
the EU Integrating project SIQS, the EU STREP EQUAM and the U.S. ARO 
contract number W91-1NF-14-1-0133.

\begin{widetext}

\appendix

\section{}
In this section, we prove a more general version of Corollary~\ref{prop:isinggroundstateln}.
A central role is played by symmetry properties of the Hamiltonian.
Notably, Hamiltonian \eqref{eq:transversefieldising} with, e.g., couplings in
Eq.~\eqref{eq:iontrapisingcouplings} or Eq.~\eqref{eq:algebraicdecayingcouplings}
is invariant if we either consider the spins from left to
right or from right to left. More formally, we denote
accordingly by $\hat I$ the transformation interchanging
spins $i \leftrightarrow N+1-i$. First however we consider
SWAP invariant Hamiltonians, i.e. invariant under interchanging
subsystems $A \leftrightarrow B$. Below Theorem~\ref{thm:groundstatelnSWAP}
we give the equivalent statement for models with $\hat I$
invariance.


Before we are in the position to state the
main theorem, we need the following simple fact:
On a bipartite system $\mathcal H_A \otimes \mathcal H_B = \mathcal H^{\otimes 2}$, let $\hat G$ be 
a SWAP-invariant Hamiltonian, i.e. $\hat S_{A\leftrightarrow B} \hat G \hat S_{A\leftrightarrow B} = \hat G$, 
then there are $b_i \in \rr$ and operators $\hat A$ and $\hat B_i$, such that
\begin{equation}
    \label{eq:swapinvariantG}
    \hat G = \id \otimes \hat A + \hat A \otimes \id + \sum_{i} b_i \hat B_i \otimes \hat B_i.
\end{equation}
To see this, note that any Hamiltonian can be written as
\begin{equation}
    \begin{split}
    \hat G &= \sum_{i,j} g_{i,j} \hat G_i \otimes \hat G_j \\
    &= \id \otimes \sum _i g_{0,i} \hat G_i + \sum _i g_{i,0} \hat G_i \otimes \id + \sum_{i,j\geq 1} g_{i,j} \hat G_i \otimes \hat G_j,
    \end{split}
\end{equation}
with $\{ \id,\hat G_1,\hat{G}_2,\dots \}$ a hermitian operator basis and $g_{i,j} \in \rr$. If 
$\hat G$ is SWAP-invariant, we have $g_{0,i}= g_{i,0}$ and the matrix g with entries $g_{i,j}$, 
$i,j\ge 1$, is real symmetric. Hence, orthogonal diagonalization $O g O^T =: \mathrm{diag}(b_i)$ 
together with the definitions
\begin{align}
    \hat A &:= \sum_i g_{0,i} \hat G_i, \\
    \hat B_j &:= \sum_j O_{i,j} \hat G_i
\end{align}
yields the desired form \eqref{eq:swapinvariantG}.

If $\hat G$ is as in \eqref{eq:swapinvariantG} with $\hat A$ and $\hat B_i$ real (but not necessarily 
hermitian) and $b_i \leq 0$ for all $i$, we call $\hat G$  \textit{negative SWAP-invariant}. Further, 
we let
\begin{equation}
    |\tilde\Phi\rangle:=\sum_{i_1,\dots,i_{N/2}}|i_1\cdots i_{N/2}\rangle|i_1\cdots i_{N/2}\rangle=:\sum_i|i,i\rangle.
\end{equation}
\begin{theorem}
\label{thm:groundstatelnSWAP}
Let $\hat G$ be a negative SWAP-invariant Hamiltonian with
a non-degenerate ground state $\ket{\psi}$. Then ground
state logarithmic negativity with respect to left half vs. right half is given by
\begin{equation}
    E_{\text{l.n.}}(\ketbra{\psi}{\psi}) = \log \tr[|\tilde\Phi\rangle\langle\tilde\Phi|\ketbra{\psi}{\psi}].
\end{equation}
\end{theorem}
\proof{
Since $\hat G$ is real and the ground state non-degenerate, we
may write (up to a global  phase)
\begin{equation}
    \ket{\psi} = \sum_{i,j} \psi_{i,j} \ket{i,j}
\end{equation}
with $\psi_{i,j}$ real. Furthermore, since $\hat G$ is SWAP invariant we have 
$\hat S_{A \leftrightarrow B}\ket{\psi} = \ket{\psi}$ such that
\begin{equation}
    \sum_{i,j} \psi_{i,j} \ket{j,i} = \sum_{i,j} \psi_{j,i} \ket{i,j},
\end{equation}
i.e., the coefficent matrix is real symmetric. Hence there is an orthogonal transformation $O$
diagonalizing $\psi$ with $\mathrm{diag(\lambda_i)} := O \psi O^T$.
This allows us to write
\begin{equation}
    \label{eq:gsschmidtdecomposition}
    \ket{\psi} = \sum_k \lambda_k \left(\sum_iO_{k,i}\ket{i}\right)\left(\sum_iO_{k,i}\ket{i}\right)
    =: \sum_k \lambda_k \ket{a_k}\ket{a_k}
\end{equation}
such that $\||\psi\rangle\langle\psi|^\Gamma\|^{1/2}_1=\sum_{i}|\lambda_i|$ and $\langle\tilde\Phi|\psi\rangle=\sum_{k} \lambda_k $.
Now, by \eqref{eq:swapinvariantG} and as $b_i\le 0$, we find for the ground-state energy
\begin{align}
    \bra{\psi} \hat G \ket{\psi} &= \sum_{i,j} \lambda_i \lambda_j \bra{a_i}\bra{a_i} \hat G \ket{a_j}\ket{a_j} \\
    &= 2\sum_{i} \lambda_i^2 \bra{a_i} \hat A \ket{a_i} + \sum_k b_k \sum_{i,j} \lambda_i \lambda_j \bra{a_i} \hat B_k \ket{a_j}^2 \\
    &\geq 2\sum_{i} \lambda_i^2 \bra{a_i} \hat A \ket{a_i}+ \sum_k b_k \sum_{i,j} |\lambda_i| |\lambda_j| \bra{a_i} \hat B_k \ket{a_j}^2 \label{eq:ineq}\\
    &= \<\tilde{\psi}| \hat G |\tilde{\psi}\>
\end{align}
with $|\tilde{\psi}\rangle := \sum_i |\lambda_i| \ket{a_i}\ket{a_i}$. As we assumed that the ground 
state is unique, we hence have $\lambda_i=\me^{\mi\phi}|\lambda_i|$ such that
\begin{equation}
    |\braket{\tilde \Phi}{\psi}|  = \sum_i |\lambda_i|=\||\psi\rangle\langle\psi|^\Gamma\|^{1/2}_1,
\end{equation}
which completes the proof.
}

The equivalence between SWAP and $\hat I$-invariance may now be exploited to obtain a result for spin 
Hamiltonians which are invariant under $\hat I$: If $\hat{H}$ is $\hat{I}$ invariant then 
$\hat{I}_B\hat{H}\hat{I}_B$ is SWAP invariant such that if $\hat{I}_B\hat{H}\hat{I}_B$ is negative SWAP 
invariant and the ground state $|\psi\rangle$ of $\hat{H}$ unique, we have
\begin{equation}
    E_{\text{l.n.}}(\ketbra{\psi}{\psi}) = \log \tr[\hat{I}_B|\tilde\Phi\rangle\langle\tilde\Phi|\hat{I}_B\ketbra{\psi}{\psi}]=\log \tr[|\Phi\rangle\langle\Phi|\ketbra{\psi}{\psi}]
\end{equation}
with $|\Phi\rangle$ as in the main text. In particular, the Ising Hamiltonian with transverse field 
in Eq.~\eqref{eq:transversefieldising} is $\hat I$-invariant if the couplings fulfil $J_{i,j} = J_{N+1-i,N+1-j}$. 
Hence, Corollary~\ref{prop:isinggroundstateln} follows from Theorem~\ref{thm:groundstatelnSWAP}. Besides 
the transverse field Ising model mentioned here, other Hamiltonians allow for similar conclusions. For 
example the ground state of the XY-model without magnetic field and arbitrary anisotropy is also determined 
by the expectation value of a single projector.

For a given coupling matrix it remains to show that it is non-positive (or non-negative) in the sense of 
Theorem~\ref{thm:groundstatelnSWAP}. In general, due to the symmetry the coupling matrix is a Hankel matrix. In the following 
we show that this is the case for algebraically decaying couplings (see Eq.~\eqref{eq:algebraicdecayingcouplings}):
\begin{equation}
    J_{i,j}= -\frac{1}{|i-j|^p},
\end{equation}
where $p \geq 0$. To simplify notation we define the $\frac{N}{2} \times \frac{N}{2}$-matrix $\mathcal{J}$ as
\begin{equation}
    \label{eq:couplings}
    \mathcal{J}_{i,j} := J_{i,N+1-j} = -\frac{1}{|N+1-i-j|^p} = -\frac{1}{(\zeta(i) +\zeta(j))^p},
\end{equation}
with $\zeta(i) = \frac{N+1}{2}-i$ . Define $d$ to be the
matrix with entries $d_{i,j} = (\zeta(i) +\zeta(j))^{-1}$.
We show that the couplings fulfil the condition
$-\mathcal{J}\geq 0$ by a result of entrywise matrix calculus.
To this end, let $f[A]$ denote the matrix obtained from $A$
by applying $f$ entrywise, i.e. $(f[A])_{i,j} := f(A_{i,j})$.
According to Theorem 1.4 in \cite{Horn69}, $\mathcal{J}$
given by \eqref{eq:couplings} is non-positive
for all powers $p \geq 0$, if $\log[d] \geq 0$ on
$D_+ = \{x\in \cc^N | \sum_i x_i = 0 \}$.
The proof of this statement follows from the fact
that there is a constant $\tau > 0$ such that
$\log[d] + \tau E \geq 0$, where the matrix $E$ is defined by $E_{i,j} = 1$.
To see this, we first rewrite $\log[d] = f[E-d]$ with $0 < 1- d_{i,j} < 1$,
where $f(x) := -\log\left(1-x\right)$.
Hence, $\log[d] + \tau E \geq 0$ iff
$f\left[E-d\right] \leq \tau E$, i.e. iff
\begin{equation}
f\left[E-d\right] \leq f[(1-\me^{-\tau}) E].
\end{equation}
This follows from two observations: (i) $f$ is a
Schur-monotone (S-monotone) function \cite{Hiai09},
i.e. for two real symmetric matrices $A,B$ with
entries $a_{i,j}, b_{i,j} \in (-1,1)$,
\begin{equation}
0 \leq A \leq B \Rightarrow f[A] \leq f[B],
\end{equation}
and (ii) $d$ is positive definite and hence we
may choose $\tau$ large enough such that
$d \geq \me^{-\tau} E$. To proof (ii), rewrite
$d$ using
\begin{equation}
d_{i,j} = \frac{1}{\zeta(i) +\zeta(j)} = \int_0^1 t^{\zeta(i) +\zeta(j)-1} \mathrm{d}t.
\end{equation}
Thus, for $x \in \cc^N$,
\begin{equation}
x^{\dag} d x = \int_0^1 \frac{1}{t} \left|\sum_i x_i t^{\zeta(i)}\right|^2 \mathrm{d}t \geq 0,
\end{equation}
with equality iff $\sum_i x_i t^{\zeta(i)} = 0$
for $t \in [0,1]$, i.e. only if $x=0$ since
all $\zeta(i)$ are distinct. Hence $d$ is
positive definite.
Thus, since $\log[d]$ coincides with $\log[d] + \tau E$ on
$D_+$, the desired positive semi-definiteness
of $\log[d]$ on $D_+$ follows.

\end{widetext}


\begin{thebibliography}{99}
\bibitem{errorFootnote} Of course, in an experimental scenario, expectation values can never be obtained
exactly and one needs to incorporate such errors into the theory.
One way is to, instead of equality, consider $|\text{tr}[\hat{C}_i\hat\varrho]-c_i|\le \epsilon_i$
along with a certain confidence and propagate these errors through the minimization,
assigning a corresponding confidence to the minimal entanglement $E_{\text{min}}$.
Most of our results however are lower bounds to $E_{\text{min}}$ that take the
form of a single expectation value $\text{tr}[\hat{C}\hat\varrho]$ such that
the usual error estimates for obtaining mean values apply.
\bibitem{SimulationParameters} In the numerical simulation we choose the specifications of trapped $^{171}$Yb$^+$ ions in a trap with vibrational frequencies $\nu_x = 2\pi\cdot 5.1$ MHz and $\nu_z = 2\pi\cdot 0.8$ MHz in transversal and longitudinal direction, respectively. Furthermore $\Omega_i = 2\pi\cdot 370$ kHz and $k = 24.03$ nm$^{-1}$.
\bibitem{hamiltonianFootnote} In fact, for ground states $|\psi\rangle$ of \eqref{eq:transversefieldising} with energy gap $\Delta$, one has, for any state $\hat\varrho$, that
$\|\hat{\varrho}^\Gamma\|_1\ge \||\psi\rangle\langle\psi|^{\Gamma}\|_1-2(2^{N}\langle\hat{H}\rangle_{\!\hat{\varrho}}/\Delta)^{1/2} $. See also Refs.~\cite{DowlingDohertyBartlett04, GuehneTothBriegel05}.
\bibitem{MPOerrorFootnote} $\|\cdot\|_F$ denotes the Frobenius norm and the minimization is over MPOs with bond dimension $D$. In Fig.~\ref{fig:lowvshighnoise} we show lower and upper bounds to this
quantity. The upper bound is obtained using a variational
compression method described in \cite{Schollwoeck11} and the lower bound may be obtained directly from the state $\hat{\varrho}(t)$ by considering its coefficient matrix: One has
$\epsilon_D(t)^2 \ge \max_l \sum_{i>D} \lambda_i^{(l)2}$, where $\lambda_i^{(l)}$ denote the operator Schmidt-coefficients of $\hat{\varrho}(t)$ with respect to the bipartition $\{1,\ldots,l\}\{l+1,\ldots,N\}$ in decreasing order.
\bibitem{Fig3Parameters} Trapping frequencies $\nu_x = 2\pi \cdot 4.7$ MHz and
$\nu_z = 2\pi \cdot 1.0$ MHz, Rabi frequency $\Omega_i = 2\pi \cdot 370$ kHz and $k = 24.03$ nm$^{-1}$.
The detuning is chosen to be $\mu = 2 \pi \cdot 10$ kHz.
\end{thebibliography}
\end{document}